\documentclass{aa}
\usepackage{graphicx}
\usepackage{subcaption}
\usepackage{txfonts}
\usepackage{hyperref}
\usepackage{xcolor}
\usepackage{csquotes}
\usepackage{siunitx}

\usepackage{longtable}
\usepackage[para]{threeparttable}
\usepackage[para]{threeparttablex}
\usepackage{placeins}
\usepackage{multicol}

\usepackage{scalerel}
\usepackage{tikz}
\usetikzlibrary{svg.path}

\definecolor{orcidlogocol}{HTML}{A6CE39}
\tikzset{
  orcidlogo/.pic={
    \fill[orcidlogocol] svg{M256,128c0,70.7-57.3,128-128,128C57.3,256,0,198.7,0,128C0,57.3,57.3,0,128,0C198.7,0,256,57.3,256,128z};
    \fill[white] svg{M86.3,186.2H70.9V79.1h15.4v48.4V186.2z}
                 svg{M108.9,79.1h41.6c39.6,0,57,28.3,57,53.6c0,27.5-21.5,53.6-56.8,53.6h-41.8V79.1z M124.3,172.4h24.5c34.9,0,42.9-26.5,42.9-39.7c0-21.5-13.7-39.7-43.7-39.7h-23.7V172.4z}electronically
                 svg{M88.7,56.8c0,5.5-4.5,10.1-10.1,10.1c-5.6,0-10.1-4.6-10.1-10.1c0-5.6,4.5-10.1,10.1-10.1C84.2,46.7,88.7,51.3,88.7,56.8z};
  }
}

\newcommand\orcidicon[1]{\href{https://orcid.org/#1}{\mbox{\scalerel*{
\begin{tikzpicture}[yscale=-1,transform shape]
\pic{orcidlogo};
\end{tikzpicture}
}{|}}}}

\begin{document}

   \title{Parsec-scale polarimetry and kinematics of a spine-sheath jet in the neutrino blazar TXS 0506+056}

   \subtitle{}
   \author{F.\,Eppel\inst{1,2,3}
          \and
          P.\,Benke\inst{4,2}
          \and
          C.\,Degli\,Agosti\inst{2}
          \and 
          J.\,L.\,Gómez\inst{5}
          \and          
          D.\,C.\,Homan\inst{6}
          \and
          M.\,Kadler\inst{1}
          \and
          Y.\,Y.\,Kovalev\inst{2}
          \and
          M.\,L.\,Lister\inst{7}
          \and
          V.\,A.\,Makeev\inst{2}
          \and
          A.\,V.\,\,Plavin\inst{8}
          \and
          A.\,B.\,Pushkarev\inst{9,10,11}
          \and
          E.\,Ros\inst{2}
          \and
          T.\,Savolainen\inst{12,13,2}
          }
   \institute{Julius-Maximilians-Universität Würzburg, Institut für Theoretische Physik und Astrophysik, Lehrstuhl für Astronomie, Emil-Fischer-Straße 31, D-97074 Würzburg, Germany\\
              \email{florian.eppel@uni-wuerzburg.de}
         \and
         Max-Planck-Institut für Radioastronomie, Auf dem Hügel 69, D-53121 Bonn, Germany
         \and
         Joint Institute for VLBI ERIC, Oude Hoogeveensedijk 4, 7991 PD Dwingeloo, The Netherlands
         \and
         GFZ Helmholtz Centre for Geosciences, Telegrafenberg, 14476 Potsdam, Germany
         \and
         Instituto de Astrofísica de Andalucía-CSIC, Glorieta de la Astronomía s/n, 18008 Granada, Spain
         \and
         Department of Physics and Astronomy, Denison University, Granville, OH 43023, USA
         \and
         Department of Physics and Astronomy, Purdue University, 525 Northwestern Avenue, West Lafayette, IN 47907, USA
         \and
         Black Hole Initiative, Harvard University, 20 Garden St, Cambridge, MA 02138, USA
         \and
         Crimean Astrophysical Observatory, 298409 Nauchny, Crimea
         \and
         Institute for Nuclear Research, 60th October Anniversary Prospect 7a, Moscow 117312, Russia
         \and
         Lebedev Physical Institute, Pushchino Radio Astronomy Observatory, Radiotelescopnaya 1a, Pushchino 142290, Russia
         \and
         Aalto University Department of Electronics and Nanoengineering, PL 15500, FI--00076 Aalto, Finland
         \and
        Aalto University Metsähovi Radio Observatory, Metsähovintie 114, FI--02540 Kylmälä, Finland
        }

   \date{Received June XX, 2026; accepted XX }

  \abstract
  {In 2017, the blazar TXS\,0506+056 showed a remarkable gamma-ray outburst simultaneously with a high-energy neutrino detection by IceCube from the same sky region. The significance of this association was found to be on the order of 3\,sigma thus providing a strong link between the neutrino emission and the blazar flare. The high-energy flare in TXS\,0506+056 was followed by a delayed radio flare, peaking in $\sim 2020$ in the aftermath of the neutrino event. We investigate the parsec-scale jet structure and dynamics of TXS\,0506+056 using 15\,GHz full-polarization VLBI observations obtained between 2009 and 2025 with the Very Long Baseline Array. Our kinematic analysis reveals moderate superluminal jet speeds of $\sim(1-2)$\,c and two quasi-stationary components. A new jet component with comparable speed was ejected contemporaneously with the 2017 IceCube neutrino event. The stacked polarization map is consistent with a stratified spine–sheath jet structure: an inner spine with EVPA aligned with the jet, surrounded by a sheath layer with perpendicular EVPA. Variability in total intensity and polarization further indicates interaction between these layers, particularly evident in characteristic linear polarization flares, associated with EVPA rotations of the quasi-stationary components. The multi-layered jet configuration is consistent with previous studies that were able to explain the neutrino emission in this TXS\,0506+056 through a spine-sheath jet structure. We suggest that TXS\,0506+056 represents an archetypal case and that similar polarization signatures and geometric light curve flares may be present in other neutrino-emitting sources, potentially offering a solution to the Doppler-crisis phenomenon observed in TeV-emitting blazars.}


\authorrunning{F.~Eppel et al.}

\maketitle

\nolinenumbers

\section{Introduction}

The origin and production mechanisms of astrophysical neutrinos are still unknown, with only a few neutrino-emitting sources identified so far. Apart from the Galactic plane \citep[][]{IceCubeGalacticPlane,2025ApJ...982...73A}, AGN are thought to contribute significantly to the astrophysical neutrino flux \citep[e.g.,][]{Plavin2020,Plavin2021,Plavin2023,Buson22,Kouch2024}. Their supermassive black holes and strong magnetic fields can act as extreme particle accelerators which are necessary for the production of high-energy neutrinos \citep[e.g.,][]{Mannheim1993,Mannheim1995}. High-energy neutrinos can be produced in many different environmental conditions and spatial locations in AGN \citep[e.g.,][]{Oikonomou22,Murase2022}. Thus, it is crucial to study the very few high-significance neutrino-emitting sources in detail to better understand the physics behind their neutrino emission. One of the most tantalizing neutrino source associations to date is the link between the IceCube neutrino track alert IC170922A and the blazar TXS\,0506+056 \citep{IceCubeTXS0506} that exhibited a strong gamma-ray flare at the time of the neutrino event. Additionally, the \cite{IceCubeTXS0506_preneutrino} identified a flare of lower energy neutrinos in 2014$-$15 coming from the same direction. TXS\,0506+056 is a TeV-emitting blazar, located at a redshift of $z=0.3365$ \citep{redshift0509}, and classified as a masquerading BL Lac object \citep{0506_masquerading}. Previous radio studies of this source have remained inconclusive about its pc-scale jet structure, e.g., \cite{Kun2019} have reported about a possible jet bend in the source and the presence of quasi-stationary features, while \cite{Ros2020} found a limb-brightened jet, suggestive of a spine-sheath structure. Additionally, the presence of a precessing helical jet structure \citep{Li2020}, a dual-jet system \citep{Britzen2019}, a supermassive binary black hole system \citep{Tjus2022,Jaroschewski2023} and gravitational lensing \citep{Britzen2025} have been claimed. In this work, we analyze the response of the radio jet after the neutrino event in 2017 and throughout a subsequent major radio flare that started rising at the time of the neutrino event. The analysis presented here and in the companion paper by \cite{0506wave} is based on the most dense and sensitive VLBI monitoring data available for this source from the MOJAVE program at 15\,GHz, carried out with the Very Long Baseline Array (VLBA). While previous studies focused on the jet structure and dynamics before and shortly after the neutrino event, we present an extensive study throughout the strong radio flare after the neutrino event and its decay. Throughout the paper, we are using a flat $\Lambda$CDM-model, with $H_0=71$\,km\,s$^{-1}$\,Mpc$^{-1}$, and $\Omega_\mathrm{m}=0.27$ \citep{Komatsu2009}. This puts TXS\,0506+056 at a luminosity distance of 1.8\,Gpc, with a linear scale of 4.8\,pc\,mas$^{-1}$, such that an apparent motion with 1\,mas\,yr$^{-1}$ corresponds to a speed of 20.8\,c.

\section{Observations and analysis}
\label{sec:obs}

\subsection{VLBI observations}

We used full-polarization VLBI images from the MOJAVE program (Monitoring of Jets in Active Galactic Nuclei with VLBA Experiments)\footnote{Data publicly available at \url{https://www.cv.nrao.edu/MOJAVE/sourcepages/0506+056.shtml}}. These data were supplemented by two epochs from a target-of opportunity (ToO) campaign recorded shortly after the neutrino event in 2017 with the VLBA; project code BR224 \citep[see][]{Ros2020}. This comprises 43 epochs in total observed between Jan~7, 2009 and Nov~23, 2025. The MOJAVE data were processed using the standard MOJAVE data reduction procedures \citep[][]{Lister2009,MOJAVE}. The typical EVPA error for MOJAVE data at 15\,GHz is about $3^\circ$ \citep{MOJAVE}. Amplitude correction coefficients were estimated and applied to all 15~GHz epochs from a comparison to the OVRO 40-m monitoring program \citep{OVRO}; consequently, the typical flux density error is 5\% \citep{MOJAVEI}. 

The ToO data were processed in a similar way using \texttt{AIPS} \citep{AIPS} for a-priori calibration and \texttt{difmap} \citep{difmap} for amplitude and phase self-calibration. Leakage solutions were derived using the \texttt{AIPS}-based \texttt{LPCAL} task \citep{LPCAL}. In order to derive the RL-phase calibration for BR224 data, i.e., the absolute calibration of the electric vector position angle (EVPA), we used the calibrator source 0528+134. For each observing session, we calculated the expected EVPA of 0528+134 by interpolating archival MOJAVE data of the source to derive an RL-phase correction solution. This solution was then applied to 0506+056 using the \texttt{RLCOR} task in \texttt{AIPS}. For the BR224 epochs we estimate an EVPA error of about $10^\circ$, based on the standard deviation of the EVPA of the calibrator source.

\subsection{Modelfitting}
\label{sec:modelfitting}

We modeled the jet structure of TXS\,0506+056 by fitting circular Gauss components to the visibility data using the \texttt{modelfit} task in \texttt{difmap}. In order to find the most consistent modelfit and to reduce human bias on the final modelfit outcome, we produced three independent modelfit versions of the entire dataset from 2009 to 2025. The different modelfit versions varied mostly in the number of fitted components and their cross-identification across epochs. We compared the different versions based on the component positions, their flux densities and position angles, and converged to a final conservative modelfit version which describes the general jet dynamics seen in all three versions best. Due to the sparse and sometimes varying uv-coverage between epochs (typically 6 scans), it is possible that during the modelfit process non-physical residual features might be fitted that cannot clearly be associated with the same feature in every epoch. 
To avoid this, we decided to use the version with the lowest number of components that still describes the data well but does not include any \enquote{orphan} components that appear only in a single epoch. This is significantly different from the approach followed, e.g., by \cite{Sumida2022} or \cite{Britzen2025} who were explicitly looking for peculiar properties in the jet and also fitted very faint components that might be of residual origin. We have verified the validity of our uv-modelfit by comparing it with other analysis methods such as image plane clustering and wavelet-based image-plane analysis \citep{WISE}, which deliver consistent results.

We calculated errors and signal-to-noise ratios (S/N) for all model components, following the method introduced by \cite{Schinzel2012}, as implemented in the \texttt{VCAT} python package \citep{VCAT}. In addition to the errors based on the S/N we have taken into account a 5\% gain error \citep{MOJAVEI}. Based on the S/N of each component, we calculated their minimum resolvable size and the brightness temperature for resolved components, following \cite{Lobanov05} and \cite{Kovalev05}.

\section{Results}
\label{sec:results}

\subsection{Jet geometry \& collimation profile}
\label{sec:geometry}

\begin{figure*}
    \centering
    \includegraphics[width=.4\linewidth]{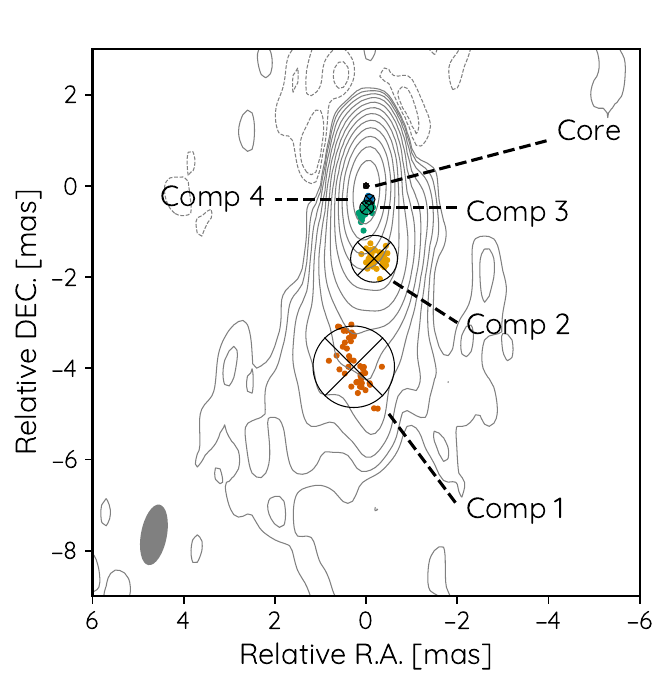}
    \includegraphics[width=.59\linewidth]{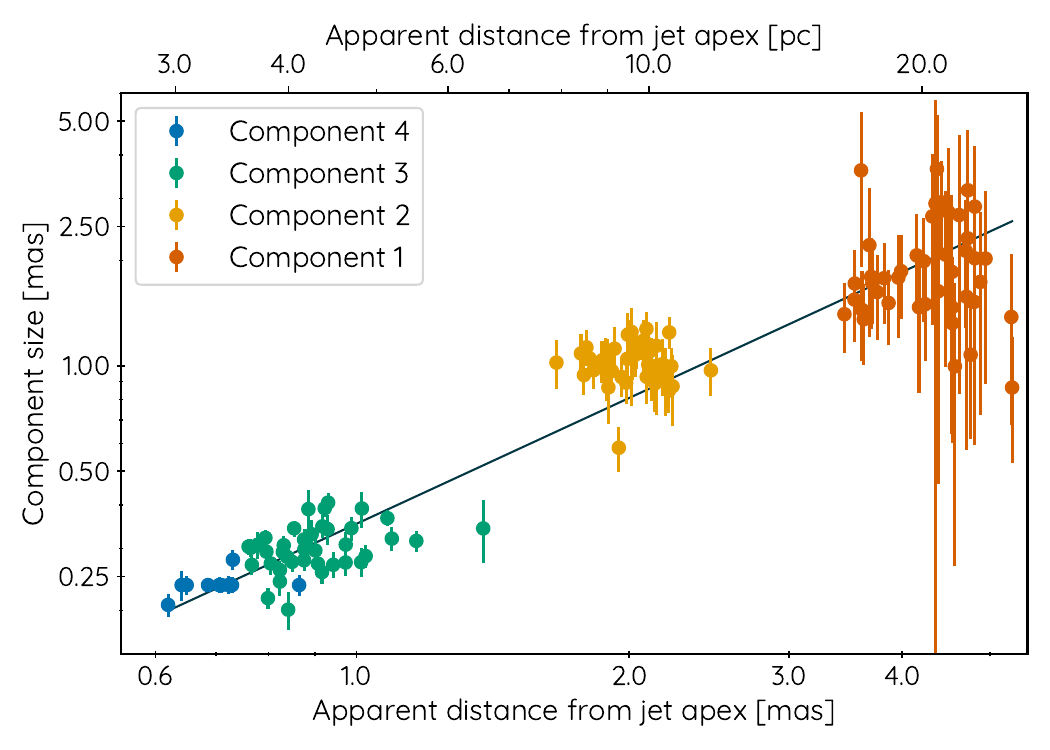}
    \caption{\textsl{Left:} Stacked total-intensity map of TXS\,0506+056, compiled from all 43 analyzed 15\,GHz VLBI images between January 2009 and November 2025. Before stacking, all images were aligned on the core component and restored with the median elliptical beam (1.35\,mas $\times$ 0.57\,mas, PA $-5.6^\circ$), displayed in the bottom left of the image. The contours start at 0.22\,mJy/beam, corresponding to five times the noise level of the stacked image and increase by factors of two. The dashed contours indicate negative intensity levels. The median modelfit components (position and size median) are plotted on top of the Stokes I contours in black, labeled with their ID. Individual component positions for every epoch are shown as dots. \textsl{Right:} Collimation profile of TXS\,0506+056 compiled from all resolved modelfit components of all 15\,GHz images.}
    \label{fig:stacked}
\end{figure*}

TXS\,0506+056 shows an extended jet structure, reaching down $\sim$7\,mas ($\sim$35\,pc) south of the core component. A stacked 15\,GHz map compiled from all 43 MOJAVE images between January 2009 and November 2025 is shown in Figure\,\ref{fig:stacked} (left). Before stacking, we convolved all images with the median beam (1.35\,mas $\times$ 0.57\,mas, PA $-5.6^\circ$) and centered them on the core component. The modelfit analysis reveals four components that we label as the core component and components 1$-$3 (jet components). Additional emission appears after 2019 between the core component and component 3 which we model as component 4 (for a detailed kinematic analysis, see Sect.\,\ref{sec:kinematics}). The median components (size and position median) are shown on top of the stacked map in Figure\,\ref{fig:stacked} (left) and their parameters are listed in Tab.\,\ref{tab:average_comps}.

\begin{table}[htbp]
\centering
\begin{threeparttable}

\caption{Median component properties as shown in Fig.\,\ref{fig:stacked} (left).}
\label{tab:average_comps}
\begin{tabular}{cccccc}
\hline\hline
\textbf{ID} & \textbf{FWHM}\tnote{a} &  \textbf{S}$^b$ & \textbf{Core Distance} & \textbf{X} & \textbf{Y} \\
           & [mas]  &    [mJy]         & [mas]                      & [mas]     & [mas]     \\
\hline
Core  & 0.11  & 642 & 0.0 & 0.0  & 0.0  \\
4$^\dagger$   & 0.24 & 556 & 0.3 & $-0.1$ & $-0.3$ \\
3  & 0.30  & 159 & 0.5 & 0.0  & $-0.5$  \\
2  & 1.03  & 91  & 1.6 & $-0.2$  & $-1.6$  \\
1  & 1.79  & 24  & 4.0 & 0.3  & $-4.0$  \\

\hline
\end{tabular}
\begin{tablenotes}
\tnote{a} Full width at half maximum, median calculated only from resolved components
\tnote{b} Flux density
\tnote{$\dagger$}~component appears in 2019
\end{tablenotes}
\end{threeparttable}
\end{table}

Following \cite{MOJAVE_opening}, we used all model components to calculate the apparent jet opening angle, $\phi_\mathrm{app}=$(28.1$\pm$4.5)$^\circ$, averaged over all observing epochs. This value is consistent with that reported by \cite{Li2020}, who used the same method on their model fit data, and by \cite{0506wave}, who used the same dataset but a method based on stacked image slices. Additionally, we used the model components to fit a collimation profile \cite[e.g.,][]{Kovalev20collimation}, assuming a power law of the form 
\begin{equation}
    w(r)=w_0(r+r_0)^k\,,
\end{equation}
where $w_0$ is the jet width 1\,mas from the jet apex, $r$ is the distance of a component to the radio core, $r_0$ the distance from the radio core to the jet apex, and $k$ the power-law/collimation index. The best fit collimation profile is shown in Fig.\,\ref{fig:stacked} (right), with $w_0=(0.354\pm0.075)$\,mas$^{1-k}$, $r_0=(0.40\pm0.17)$\,mas, and $k=1.19\pm0.15$. The jet profile is consistent with a conical jet (i.e., $k\sim1$). The distance to the jet apex, $r_0$, is consistent with the upper limit of the core shift analysis discussed by \cite{0506wave}. Though, \cite{0506wave} argue that the core shift is likely much closer to zero, suggesting that the assumption of a simple power law profile might be an oversimplification. A better description of the jet profile might be possible using a broken power law \citep[cf.][]{Kovalev20collimation,Burd2022}. However, the exact location of the break point is difficult to constrain, since there are only very few model-fit components between 1$-$2\,mas from the core. Moreover, we note that the collimation profile and opening angle estimates may depend on the adopted modelfit, since different plausible component configurations can sample the transverse jet structure differently, introducing additional systematic uncertainties not captured by the formal fit errors presented here.

\subsection{Jet kinematic analysis \& wave propagation}
\label{sec:kinematics}

\subsubsection{Component dynamics}
\label{sec:dynamics}

\begin{figure*}
    \centering
    \includegraphics[width=\linewidth]{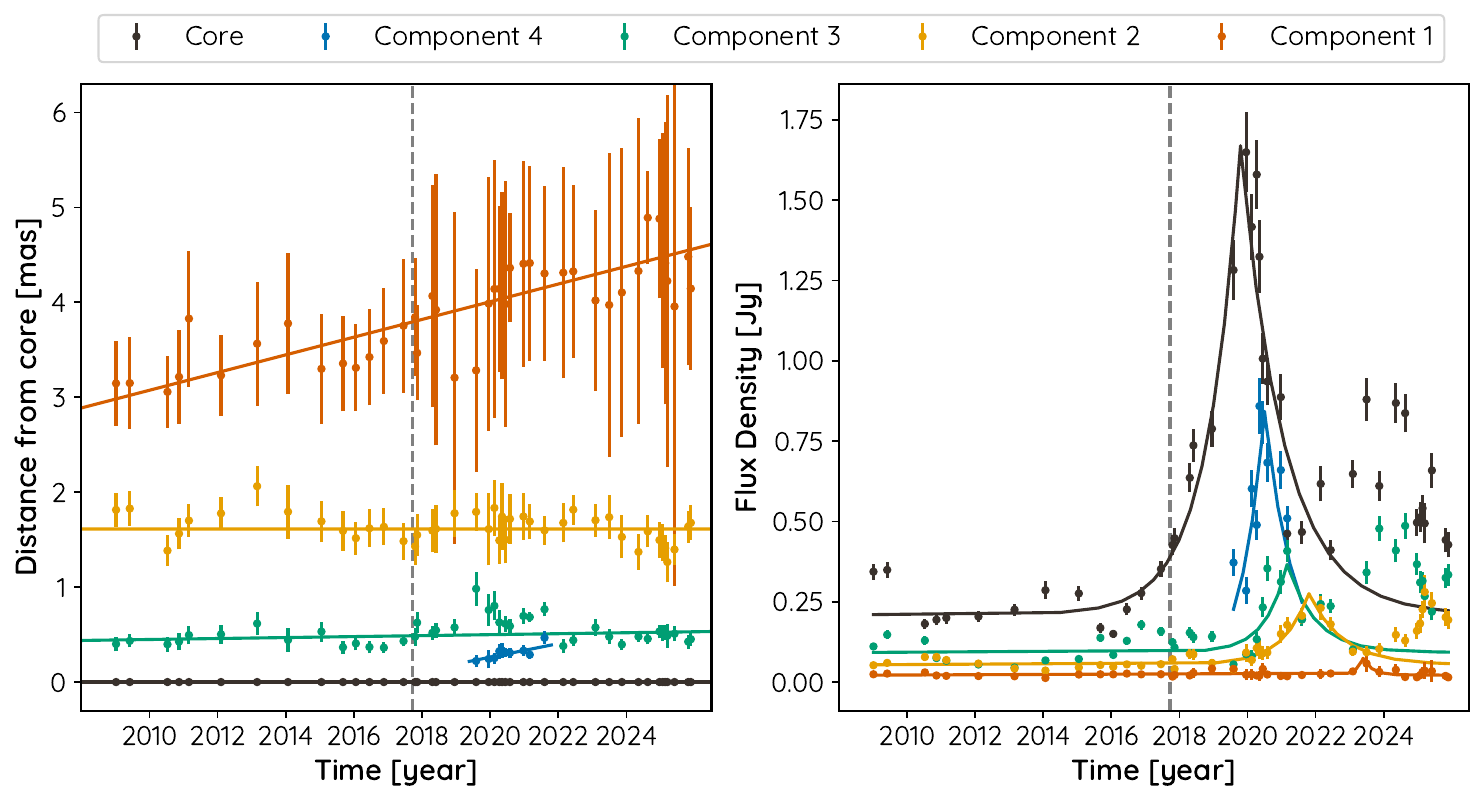}
    \caption{\textsl{Left:} Core separation vs. time for all modelfit components found in TXS\,0506+056. Components 2 and 3 are stationary while components 1 and 4 move at slightly superluminal speeds of 1$-$2\,c. Component 4 appears $\sim2$\,years after the neutrino alert IC170922A (grey dashed line). Its ejection time is consistent with the neutrino event. \textsl{Right:} Component light curves with exponential flare fits as described in Eq.\,\eqref{eq:flare}. The flare peak times show a significant delay between the components, with components further downstream flaring at later times, consistent with the flux density wave propagation reported by \cite{0506wave}. 
    Here, the fits characterize the kinematic and variability properties of the individual components.}
    \label{fig:kinematic_15ghz}
\end{figure*}

We investigated the jet dynamics by analyzing the model components and their distance to the core component versus time. A kinematic plot for all components is shown in Fig.\,\ref{fig:kinematic_15ghz} (left). We modeled the core distance versus time with a linear fit for all components. The fit results are presented in Tab.\,\ref{tab:kinematic_fit}.
The kinematic plot clearly reveals that components 3 and 2 do not show significant motion relative to the core component. 
Component 1 is moving outwards at a mildly relativistic speed of $(1.93\pm0.45)$\,c, slightly faster than previously reported component motion in this source \citep[see][]{Lister2021}. After $\sim$mid 2019, a new component appears close to the core (component 4) whose motion cannot be followed robustly due to the stationary feature at around 0.5 mas south of the core (i.e., component 3). A linear fit to its motion indicates mildly relativistic motion at speeds of $\sim(1-2)$\,c, similar to component 1 and previous reports by \cite{Lister2021}. 
We note that the fitted ejection time of component 4 from the 15\,GHz core, $t_0=(2016.3\pm2.1)$\,yr, is consistent with the neutrino arrival time observed by IceCube. 

For each component, we calculated the critical Doppler factor, $\delta_\mathrm{crit}=\sqrt{1+\beta_\mathrm{app}^2}$. This corresponds to the Doppler factor at the critical viewing angle, i.e., the angle which maximizes the apparent jet speed (see Tab.\,\ref{tab:kinematic_fit}). Since components 2 and 3 are stationary, no sensible ejection epoch and critical Doppler factor could be determined for them.

\begin{table}[htbp]
\centering
\begin{threeparttable}

\caption{Kinematic fit results as shown in Fig.\,\ref{fig:kinematic_15ghz} (left).}
\label{tab:kinematic_fit}

\begin{tabular}{ccccc}
\hline\hline
ID & $\mu_\mathrm{app}^a$ & $\beta_\mathrm{app}^b$ & $t_0^c$ & $\delta_\mathrm{crit}^d$ \\
 & [$\mu$as/yr] & [c] & [yr] &   \\
\hline
4 & 72 ± 35 & 1.48 ± 0.73 & 2016.3 ± 2.1 & 1.84 ± 0.64 \\
3 & 5.1 ± 2.5 & 0.105 ± 0.052 & - & - \\
2 & 0.0 ± 5.7 & 0.00 ± 0.12 & - & - \\
1 & 93 ± 22 & 1.93 ± 0.45 & 1977.0 ± 9.2 & 2.24 ± 0.42 \\

\hline
\end{tabular}
\begin{tablenotes}
\tnote{a} Angular speed
\tnote{b} Apparent speed
\tnote{c} Ejection epoch under assumption of a constant component speed
\tnote{d} Doppler factor at the critical viewing angle
\end{tablenotes}
\end{threeparttable}
\end{table}

\begin{table*}[htbp]
\centering
\begin{threeparttable}

\caption{Fit parameters for individual component flares modeled by Eq.\,\eqref{eq:flare} and shown in Fig.\,\ref{fig:kinematic_15ghz} (right).}
\label{tab:0506_variability_doppler}

\begin{tabular}{ccccc}
\hline\hline
ID & $\Delta S_\mathrm{max}$ & $t_\mathrm{max}$ & $\tau$ & $S_0$ \\
 & [Jy] & [years] & [years] & [Jy] \\
\hline
Core & $1.463 \pm 0.065$ & $2019.792 \pm 0.031$ & $0.979 \pm 0.044$ & $0.2106 \pm 0.0060$ \\
4 & $0.84 \pm 0.16$ & $2020.499 \pm 0.026$ & $0.69 \pm 0.31$ & $0.00 \pm 0.19$  \\
3 & $0.274 \pm 0.028$ & $2021.164 \pm 0.052$ & $0.640 \pm 0.075$ & $0.0930 \pm 0.0027$ \\
2 & $0.223 \pm 0.031$ & $2021.796 \pm 0.069$ & $0.76 \pm 0.11$ & $0.0545 \pm 0.0023$  \\
1 & $0.053 \pm 0.071$ & $2023.40 \pm 0.19$ & $0.20 \pm 0.26$ & $0.0222 \pm 0.0014$ \\

\hline
\end{tabular}
\begin{tablenotes}
\end{tablenotes}
\end{threeparttable}
\end{table*}

We find that the peak brightness temperature of the core, $T_\mathrm{b}=(1.36\pm0.12)\times10^{12}$\,K, was observed on August 4, 2019, shortly before the core reached its maximum flux density. This value exceeds the equipartition limit significantly \citep[$T_\mathrm{b,eq}=5\times 10^{10}$\,K,][]{Readhead}. While AGN can reach intrinsic brightness temperatures of several times $10^{11}$\,K during flaring episodes \citep[e.g.,][]{Homan2006}, the exceptionally high brightness temperature reached by the core exceeds these values and therefore indicates that strong Doppler boosting is likely contributing.

\subsubsection{Component light curves}
\label{sec:lightcurves}

The light curves of all modelfit components are shown in Fig.\,\ref{fig:kinematic_15ghz}. All components exhibit a prominent flare after $\sim 2020$. We modeled the flares of the individual component light curves following the approach introduced by \cite{Lahtenmaki1999PaperIII}, with exponential flares of the form
\begin{equation}
\label{eq:flare}
    S(t)= \begin{cases}
\Delta S_\mathrm{max}\mathrm{e}^{(t-t_\mathrm{max})/\tau} + S_0 & \text{if } t \leq t_\mathrm{max} \\
\Delta S_\mathrm{max}\mathrm{e}^{(t_\mathrm{max}-t)/1.3\tau} +S_0 & \text{if } t > t_\mathrm{max}
\end{cases},
\end{equation}
where $\Delta S_\mathrm{max}$ is the flare amplitude above a rest flux density, $S_0$, $t_\mathrm{max}$ is the peak time of the flare, and $\tau$ the variability time scale. The resulting fits are displayed in Fig.\,\ref{fig:kinematic_15ghz} (right) and the fit parameters are listed in Tab.\,\ref{tab:0506_variability_doppler}. For the core component and components 2$-$4, we limited the fit to data taken before 2024, since there is additional activity after that time that is likely unrelated to the major flare. For component 1, the full time range was used but we detect only a small flux peak with large uncertainties on its amplitude and peak time.

It is evident from Fig.\,\ref{fig:kinematic_15ghz}, that the flux density maxima of components further downstream are shifted to later times. As discussed in detail by \cite{0506wave}, if this disturbance is associated with the same plasma propagating downstream, it suggests a significantly higher apparent speed than what is observed from the component motion.\footnote{As discussed by \cite{0506wave}, using the flare peak times of the components from Tab.\,\ref{tab:0506_variability_doppler} and their position at that time, a propagation speed of $(14.8\pm1.9)$\,c can be inferred. An independent estimation based on image slices suggests $(20.9 \pm 0.9)$\,c \citep[see][]{0506wave}}.

We note that the variability timescale suggested by the flare fits does not change significantly between the components, while their size varies drastically from $\sim0.1$\,mas (core component) to $\sim1$\,mas (component 2). In many publications \citep[e.g.,][]{Jorstad05,Jorstad17}, it is assumed that the variability timescale corresponds to the light-travel time across the component and that the flux decay is driven by radiative losses, which can be used to estimate variability Doppler factors, $\delta_\mathrm{var}$. Since the variability timescale for the components in TXS\,0506+056 is almost constant, this assumption would lead to relatively moderate Doppler factors $\sim4$ at the core, increasing further downstream up to $\sim50$ for component 2. This strong increase of $\delta_\mathrm{var}$ with increasing distance from the core would require a significant acceleration of the flux density wave or a changing viewing angle in the jet. However, based on the wave propagation analysis by \citep{0506wave} there is no significant evidence for such scenarios. Thus, the underlying assumptions that the variability timescale corresponds to the light-crossing time and that the true geometry of the components is similar to a uniform face-on disk are likely not fulfilled for this source, e.g., due to the presence of a stratified jet \citep[see][]{0506wave}.

\subsubsection{Modelfit polarization \& EVPA swing}
\label{sec:polarization_evpa_swing}

\begin{figure*}
    \centering
    \includegraphics[width=.85\linewidth]{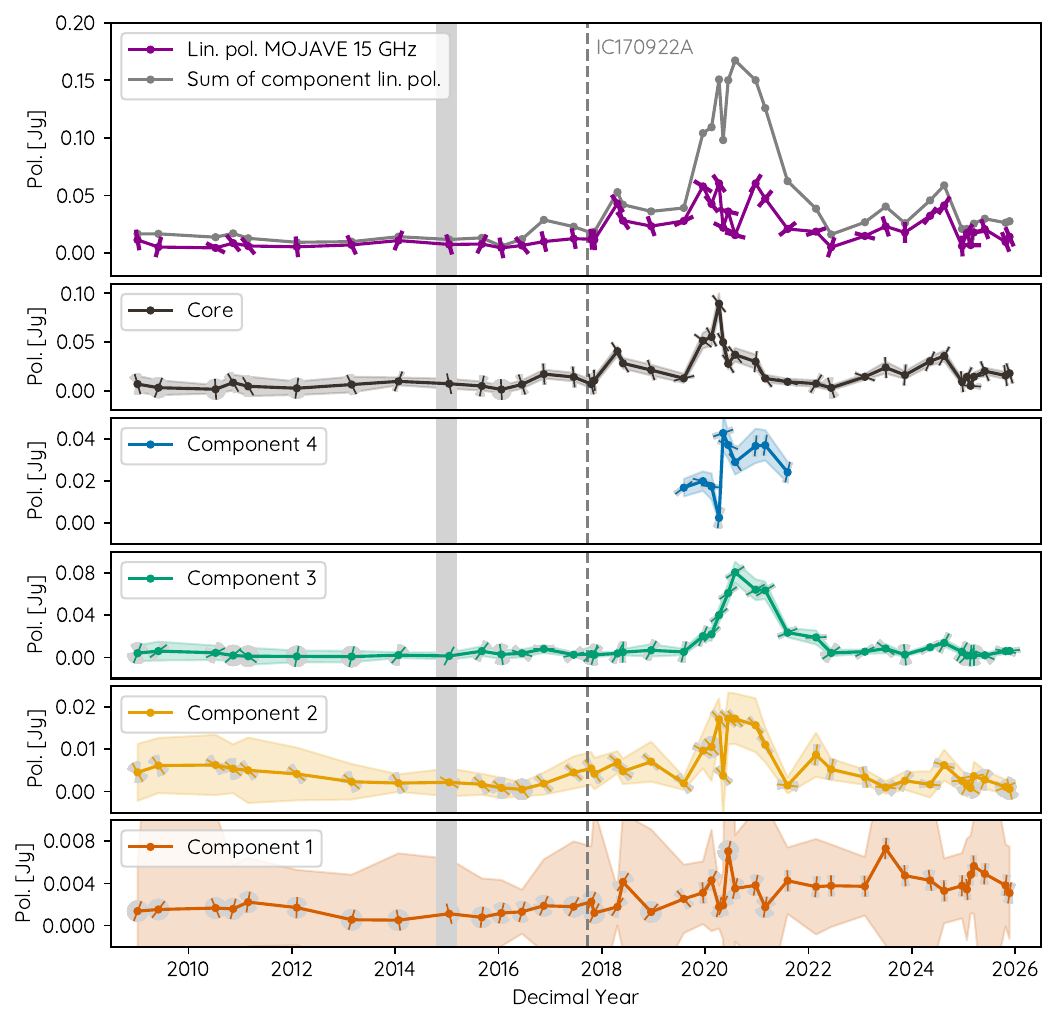}
    \caption{Light curves of linear-polarized intensity of the integrated image (top panel) and individual components (bottom five panels). The EVPA direction is indicated by the inclined lines plotted on top of the data points and its uncertainty indicated by the shaded gray area. The top panel additionally shows the algebraic sum of the linear polarization values of all components (gray dots). The time of IC170922A is highlighted by the dashed line and the time of the low-energy neutrino flare is highlighted by the gray area.}
    \label{fig:pol_lcs}
\end{figure*}

For all model components, we derived linearly polarized flux densities and EVPA values by loading the Stokes-I models for every epoch back into \texttt{difmap} and performing a modelfit in Stokes Q and U while keeping the component position and size fixed, i.e., only fitting the Stokes Q and U flux densities of the components. The resulting component light curves in linearly polarized intensity are shown in Fig.\,\ref{fig:pol_lcs} (bottom 5 panels), with their corresponding EVPA direction indicated by the tilted lines plotted on top of the data points. The top panel of Fig.\,\ref{fig:pol_lcs} displays the integrated linear polarization of the clean model (purple curve) and the summed linear polarization from all individual components (gray curve). Uncertainties for the Stokes Q and U flux densities of the components were calculated with the same method that was used for Stokes I, following \cite{Schinzel2012}, but using the Stokes Q and U residual maps to determine the respective component S/N in Q and U. An automated implementation of this was made available as part of the \texttt{VCAT} python package \citep{VCAT}. 

The integrated polarization flux density (Fig.\,\ref{fig:pol_lcs}, upper panel, purple curve) starts rising within a few months after the neutrino event in 2017. During the time of the major total intensity flare peaking in $\sim2020$, it is rising even further for a short time and then suddenly drops. At the same time, the algebraic sum of the individual component polarizations (Fig.\,\ref{fig:pol_lcs}, upper panel, gray curve) rises even further, similar to the behavior in total intensity. This indicates that the total amount of linear polarization in the jet is significantly higher throughout 2020$-$2022 (i.e., during the wave propagation) than in previous states of the source. However, since different regions (i.e., components) of the jet have different EVPAs, a polarization flare of multiple components at the same time causes a net depolarization which is reflected by the purple curve showing a minimum at that time. Throughout the decay of the major total intensity flare the purple curve shows a second peak, again followed by a steep drop associated with an EVPA rotation by $\sim 90^\circ$. Afterwards, in 2023$-$2024 it returns to a similar state as before the neutrino event. This polarization behavior is similar to the one observed for the neutrino-candidate blazar PKS\,0215+015 \citep{Eppel0215} and could be explained by the interaction of polarized features with different speeds and EVPAs.

When looking at the polarization light curves of the individual components, it can be seen that the core component flares first. Shortly after, component 4, located $\sim0.3$\,mas away from the core, dominates the polarized emission under significant rotation of its EVPA. Only a few weeks later, component 3 ($\sim0.5$\,mas away from the core) shows its polarization flare peak, also under a significant EVPA rotation as compared to values observed before 2020. The time delay of the polarization flares between the core component, component 4, and component 3 is consistent with the wave propagation seen in total intensity. The polarization detection of components 2 and 1 is much less significant. Still, component 2 shows a mild flare shortly after 2020 which happens before the total intensity wave passes through this feature. However, an EVPA rotation of component 2 is observed only in 2022, which would again be consistent with the time when the total intensity wave passes through component 2.

Upon closer inspection of Fig.\,\ref{fig:pol_lcs} it can be seen that also after the low-energy neutrino flare found in 2014$-$15 by the \cite{IceCubeTXS0506_preneutrino}, component 3 shows an EVPA rotation, starting in mid 2016. This could possibly indicate increased jet activity already previous to the 2017 IceCube neutrino, as further discussed in App.\,\ref{app:additional_wave}.

\subsection{Spine-Sheath Morphology}

\begin{figure*}
    \centering
    \includegraphics[width=\textwidth]{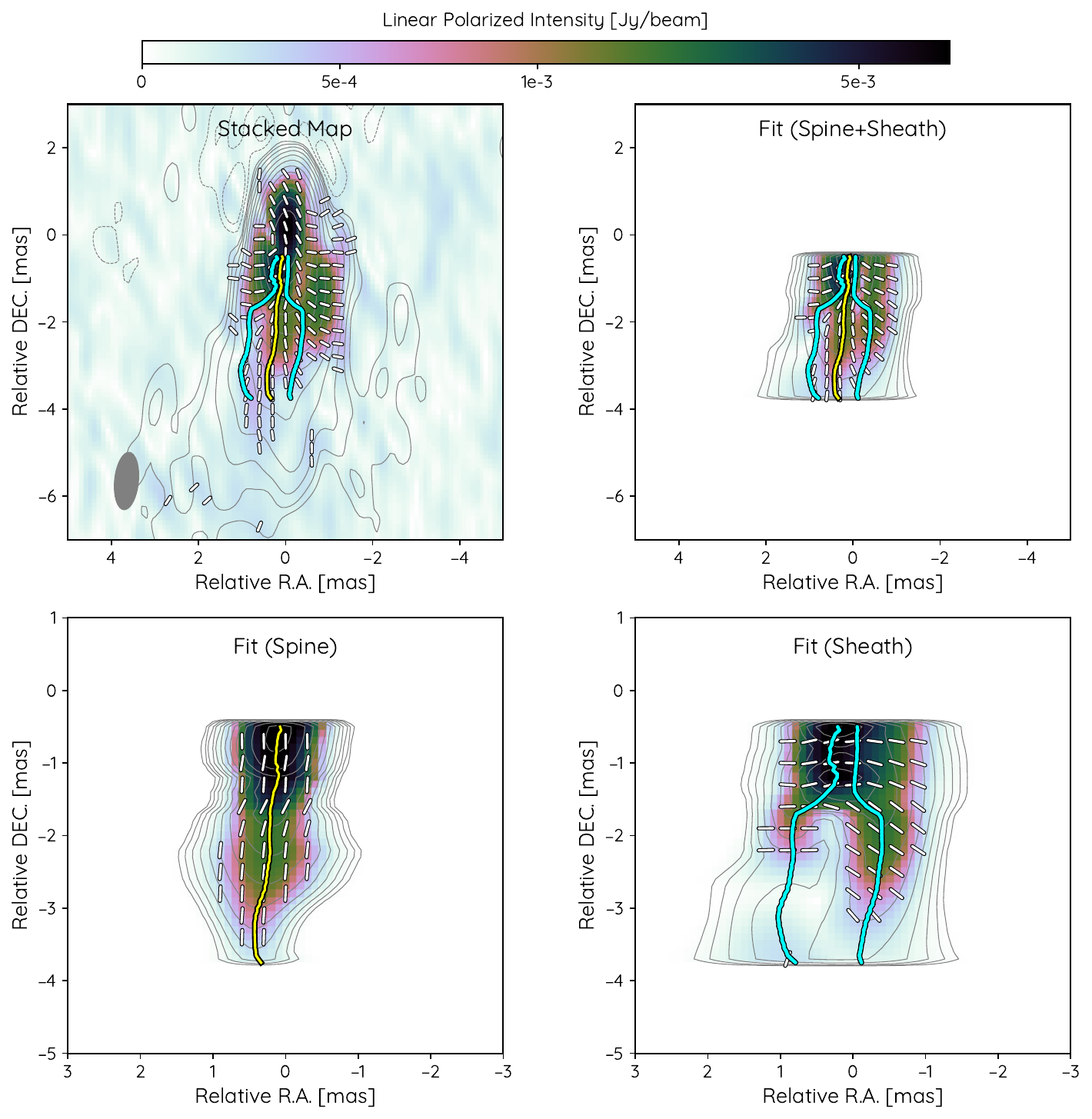}
    \caption{Stacked 15\,GHz polarization image compiled from all 43 VLBI observations from January 2009 to November 2025 (top left). The contours correspond to total intensity (cf. Fig.\,\ref{fig:stacked}, left) and the color map displays linear polarized intensity with the EVPA direction indicated by the white lines. The structure was modeled with a spine-sheath configuration in full-polarization. The center position of the different jet layers is overplotted in cyan (sheath) and yellow (spine). The spine-sheath model image is shown in the top right plot. Separate spine and sheath plots are shown in the two bottom panels.}
    \label{fig:spine_sheath}
\end{figure*}

The presence of slow jet components (see Sect.\,\ref{sec:kinematics}) and a significantly faster flux density wave at the same time as discussed by \cite{0506wave} suggest that the jet of TXS\,0506+056 is likely composed of multiple layers. This has been suggested for several other neutrino-candidate blazars in the literature \citep[e.g.,][]{Eppel0215,Kim0735,Sauron}. As a consequence, a simple model consisting of only circular Gauss components cannot fully describe the real structure of the jet. More detailed insight into the jet structure of TXS\,0506+056 can be gained by using polarization information. However, due to the relatively low fractional polarization of the source ($\lesssim5$\%), the dynamic range of the polarization maps is significantly lower than for Stokes I. In order to improve the dynamic range and to get the most sensitive polarization image of TXS\,0506+056, we stacked the data from all VLBI observations, by using the same stacking approach as discussed in Sect.\,\ref{sec:geometry}, but now also stacking the Stokes Q and Stokes U maps before combining them to a stacked linear polarization and EVPA map. The resulting full polarization stacked map is shown in Fig.\,\ref{fig:spine_sheath} (top left). This map clearly reveals that the jet consists of differently polarized regions. While the outer jet layers exhibit an almost horizontal EVPA (i.e., $\chi\sim90^\circ$), the EVPA in the central region and in the core region is almost perpendicular to that (i.e., $\chi\sim0^\circ$). This structure resembles the spine-sheath scenario, previously observed in other sources, such as 4C\,+01.28 \citep{Attridge1999} and other jetted AGN \citep[e.g.,][]{Pushkarev2005,Gabudzda2014}. We note that the polarization pattern could also be caused by a helical or toroidal magnetic field \citep[e.g.,][]{Gabudzda2014,Kramer2021}. Moreover, it shows some qualitative similarities to features reported in other compact AGN jets with very small viewing angles \citep{Sauron}, particularly in terms of EVPA patterns and localized depolarization near the core. However, polarization morphology alone is not a unique indicator of jet orientation and can arise in a range of physical scenarios, including stratified jet structures.

\begin{figure*}
    \centering
    \includegraphics[width=\linewidth]{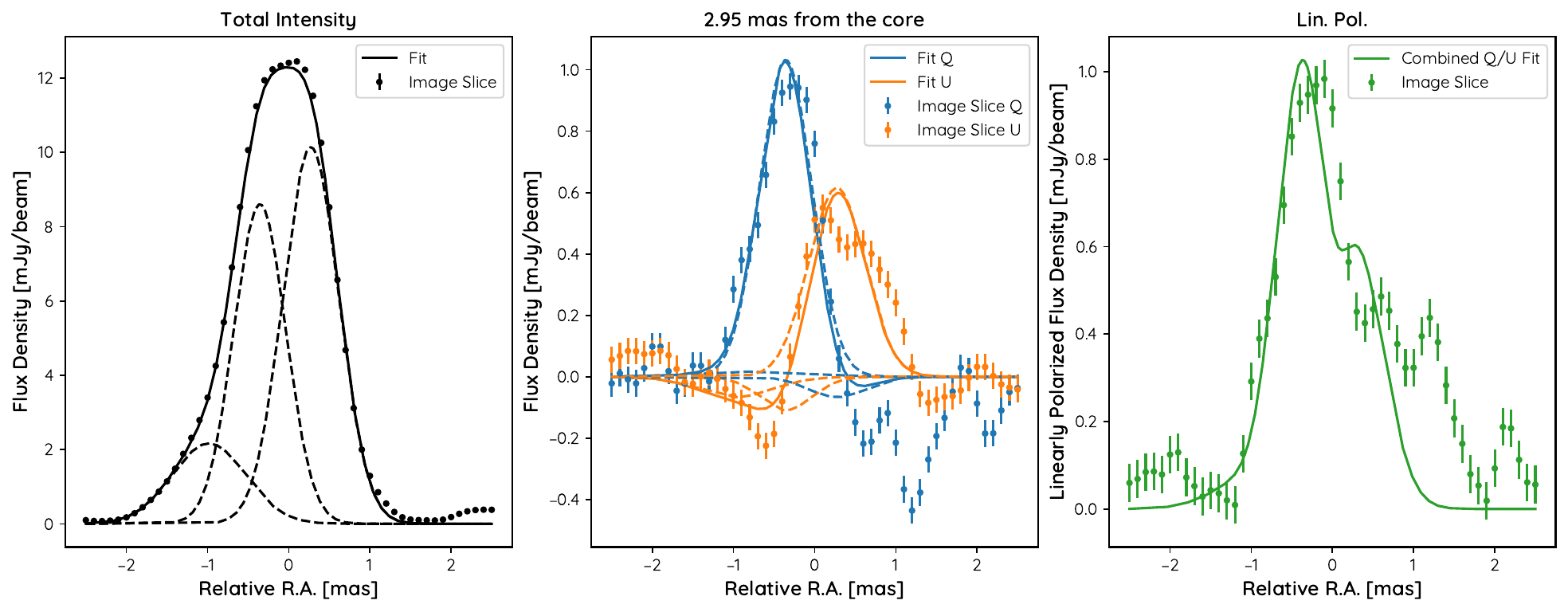}
    \caption{Exemplary slice fit in total intensity (left), Stokes Q and U (center), as well as linear polarized intensity (right), taken from the stacked image of TXS\,0506+056, 2.95\,mas South of the core component. The data points with error bars indicate the image slice data, while the dashed lines depict the individual Gauss fits of spine and sheath layers. The solid lines indicate the sum of spine- and sheath-fits.}
    \label{fig:spine_sheath_slice}
\end{figure*}

In order to characterize this structure better and to test if it can be explained by a spine-sheath configuration, a similar approach as described by \cite{Murphy2013} was followed to model the structure in full polarization. Based on the stacked map shown in Fig.\,\ref{fig:spine_sheath} (top left), horizontal slices across the jet width were taken in full-polarization (i.e., Stokes I, Q, and U). Every slice was modeled by the sum of three Gaussians. The choice of three Gaussians is motivated by the spine-sheath model \citep[e.g.,][]{Tavecchio2014,Tavecchio2015}, where the central Gaussian represents the spine, surrounded by two sheath-layers. One sheath-layer is positioned to the right of the spine, the second one is positioned left of the spine, both in equal distance $\Delta x$ to the spine. An example slice at 2.95 mas away from the core, with the corresponding best fit Gaussians is shown in Fig.\,\ref{fig:spine_sheath_slice}. The left panel shows the slice in total intensity, the center panel shows Stokes Q and U, and the right panel the total linear polarization calculated by combining the Q and U slice. The slice values from the stacked image are depicted as dots, with error bars corresponding to the image noise of the stacked image (i.e., $\sigma_\mathrm{rms}=56\,\mu$Jy/beam). Each of the three fitted Gaussians shares its width $\sigma_i$ and position $x_{0,i}$ across polarizations, while the Stokes Q and U amplitudes $Q_i$ and $U_i$, as well as the fractional polarization, $m_i$, are independent parameters. In general, every slice is described by
\begin{equation}
    \begin{pmatrix} I(x) \\ Q(x) \\ U(x) \end{pmatrix} =  \sum_{i=0}^3 \begin{pmatrix} I_i \\ Q_i \\ U_i \end{pmatrix} \mathrm{exp}\left(-\frac{1}{2}\left(\frac{x-x_{0,i}}{\sigma_i}\right)^2\right),
\end{equation}
with $I_i=\frac{\sqrt{Q_i^2+U_i^2}}{m_i}$. The fractional polarization is constrained to $0<m_i<1$ by definition. The position of the spine, $x_\mathrm{0,spine}$, is a free parameter in the fit, while the positions of the sheath-layers are determined by $x_\mathrm{0,sheath,r/l}=x_\mathrm{0,spine}\pm\Delta x$, and $\Delta x$ determined by the fit. In total, the fit includes 14 free parameters for every slice (i.e., $m_i,Q_i,U_i,\sigma_i$ for every Gaussian, plus $x_\mathrm{0,spine}$ and $\Delta x$). Since the parameter space with 14 free parameters is highly degenerate, a nested sampling algorithm \citep[e.g.,][]{NestedSampling} was used to explore the posterior distribution for every slice and to find the best fit parameters. The fitting was implemented in \texttt{python} with the \texttt{dynesty} package \citep{dynesty}, which is a dynamic nested sampling implementation that adapts the sampling strategy during runtime to focus on regions of high likelihood and improve convergence in complex parameter spaces. Within \texttt{dynesty}, the random walk (\enquote{rwalk}) sampling method was used, with 500 live points and a stopping criterion of dlogz = 0.01, which sets the convergence threshold for terminating the nested sampling procedure. Slices were taken starting at 0.5\,mas distance South of the core, every 0.05\,mas, and up to a distance of 3.8\,mas from the core. The Gauss fits were performed for every slice individually. However, in order to ensure smoothness between the slices, the center position $x_\mathrm{0,spine}$ and the FWHM were constrained to only change by $<0.1$\,mas as compared to the previous slice, as well as $\Delta x$ to not change by more than 20\% from one slice to the other. Additionally, the minimum FWHM for every Gaussian was set to 0.3\,mas, which is on the order of the beam width.

The obtained spine-sheath fit for all slices combined is shown in Fig.\,\ref{fig:spine_sheath}. In the top left panel, the center positions of the Gauss fit for the spine (yellow line) and the sheath layers (cyan lines) are shown, on top of the stacked map. The top right panel shows the combined spine-sheath model in full-polarization, which resembles the structure seen in the stacked map. The bottom two panels show the model, zoomed in, and separated into spine (bottom left) and sheath (bottom right). Here, it is clearly visible that the EVPA direction of the narrower spine is aligned with the jet direction, while the wider sheath exhibits an EVPA perpendicular to the jet direction. Since the slice fits depend on the previous slice, due to the introduced smoothness condition, the fit can potentially be influenced by the selection of the start slice (whose fit is less constrained). In order to test the severity of this effect, several alternative fits were performed with starting points between (0.2$-$1.0)\,mas and slightly modifying the smoothness conditions. Starting even further upstream than 0.2\,mas from the core is not feasible, since in this region, the spine and sheath structures cannot clearly be distinguished from each other, as they are blurred by the beam. All different fit versions are able to reproduce the polarization structure observed in the stacked map, but they differ by the amount of flux density fitted into the spine or the sheath. However, all fits consistently show that there is a wider layer with horizontal EVPA and a narrower structure in the center, with EVPAs aligned with the jet direction, as represented by the fit displayed in Fig.\,\ref{fig:spine_sheath}. We note that this transverse structure refers to the projected morphology in the plane of the sky and is therefore affected by beam convolution and projection effects, such that an estimate of the intrinsic spine-to-sheath thickness ratio would require additional assumptions on the jet geometry and viewing angle and is not robustly constrained by the present data. Moreover, the apparent sudden broadening of the sheath layers between 1$-$2\,mas from the core is likely not a robust feature, as the transition coincides with the edge of the central beam and the upstream part may therefore be affected by limited resolution. If physical, however, this behavior could indicate a change in jet collimation, as briefly discussed in Sect.\,\ref{sec:geometry}.

At this point it is important to keep in mind that the spine-sheath fitting described in this work is based on a stacked map that was compiled from more than 15\,years of observations. As a consequence, it is not possible to derive dynamics from this analysis or to identify which layer of the jet corresponds to the slow motion, as described in the component kinematics (see Sect.\,\ref{sec:kinematics}), and which one corresponds to the fast bulk motion seen as the fast wave propagation \citep{0506wave}. However, the presented spine-sheath fit provides an additional confirmation that the jet in TXS\,0506+056 consists of multiple layers, consistent with what was reported in the previous sections and the discussion by \cite{0506wave}. We note that \cite{0506wave} argue that the spine is relatively radio-faint, while most of the radio-emission originates from the sheath. At first, this might seem contradictory to the fits shown in Fig.\,\ref{fig:spine_sheath}, where spine and sheath are about equally bright in total intensity. However, what is fit as the spine in the stacked map is likely the result of an interaction between spine and sheath layers during passage of the wave. In the picture presented by \cite{0506wave}, this interaction naturally happens between the outer part of the spine and the inner part of the sheath. As a result, the total intensity emission might be episodically dominated by emission from the spine, coming from its outer edges, while the majority of the total intensity emission typically originates from the sheath. Therefore, in the stacked map, we see emission from both layers that likely is a result of an interaction between them, integrated over 17 years of observation. An additional tracer of the interplay of spine and sheath could be the EVPA rotations observed during the wave propagation (see Sect.\,\ref{sec:polarization_evpa_swing}), which could simply be caused by variations in the relative dominance of the spine vs. sheath emission.
Due to the limited dynamic range of most individual images it is challenging to investigate the dynamics of the spine and sheath layers and their interaction further with the available data. In the future this could become possible with higher sensitivity polarization observations, e.g., with upcoming next-generation facilities such as the SKA \citep{SKA} or the ngVLA \citep{ngVLA}.

\section{Discussion}
\label{sec:discussion}

\subsection{Kinematics in light of the wave propagation}

\begin{figure}
    \centering
    \includegraphics[width=\linewidth]{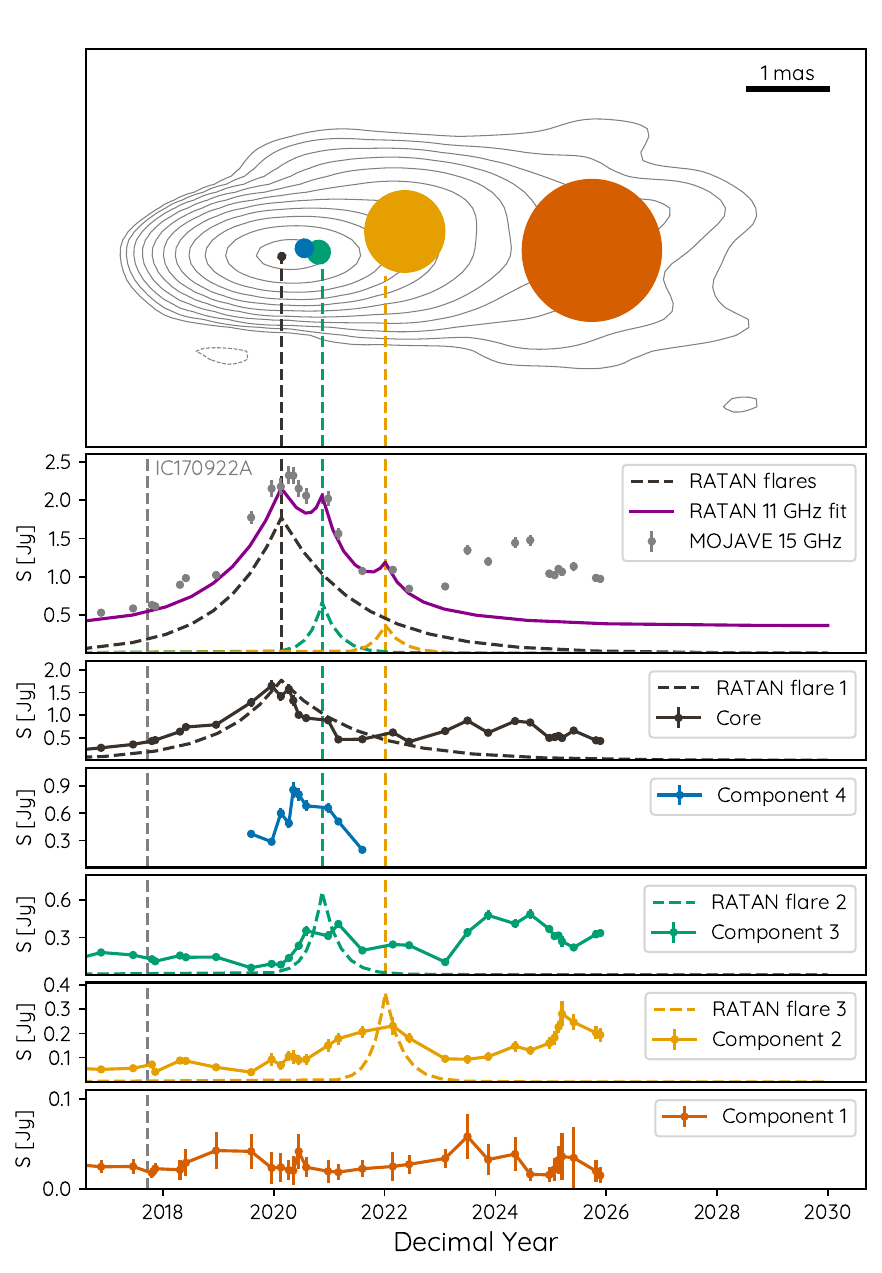}
    \caption{\textsl{Top panel:} Stacked total intensity contours of TXS\,0506+056 with median components from Tab.\,\ref{tab:average_comps} overplotted and rotated by 95\,$^\circ$ (counterclockwise) for better visualization. \textsl{Bottom panels:} Light curves of individual components (bottom five panels) and integrated total intensity light curve. The RATAN 11\,GHz fit and flare decomposition is taken from \cite{Baikal0506}. The peak time of the first flare was aligned with the core position in the upper panel, other times are mapped linearly by assuming the wave speed based on modelfit light curves \citep[$0.71$\,mas/yr, see][]{0506wave}.}
    \label{fig:waveplot}
\end{figure}

As discussed by \cite{0506wave} the jet of TXS\,0506+056 likely consists of at least two different layers, propagating at significantly different speeds. The stacked polarization map (see Fig.\,\ref{fig:spine_sheath}) supports the presence of a stratified spine-sheath jet, consistent with the scenario suggested by \cite{0506wave} and previous findings by \cite{Ros2020}. We see clear evidence of slow moving jet components with a speed of $\sim(1-2)$\,c from the kinematic analysis (see Sect.\,\ref{sec:kinematics}) and, at the same time, significantly faster plasma in form of a fast flux density wave propagating through the jet \citep{0506wave}. In agreement with the discussion by \cite{0506wave}, each Gaussian component could represent a stationary recollimation shock that is undergoing a shock-shock interaction \citep[e.g.,][]{Fromm2013Kinematic} with a faster spine-feature, resulting in the observed wave-pattern in the component light curves. A similar geometrical flare explanation was previously discussed to happen at much smaller jet scales, in order to explain X-ray and gamma-ray flare patterns possibly related to neutrino-production \citep[e.g.,][]{Hervet2019,Novikova2023,Blinov2025}. 

We note that the individual component flares of the wave propagation even seem to be present in the total intensity light curve. To better visualize this, Fig.\,\ref{fig:waveplot} shows the MOJAVE 15\,GHz total intensity light curve, as well as the individual component light curves for comparison. Moreover, the flare decomposition of RATAN-600 total intensity data as published by \cite{Baikal0506} is plotted on top of the light curves. For better visualization, a rotated stacked map of TXS\,0506+056 is plotted on top of Fig.\,\ref{fig:waveplot}, including the median modelfit components as listed in Tab.\,\ref{tab:average_comps}. The peak time of the first flare was aligned with the VLBI core position (core component), and other times were mapped linearly with the wave velocity based on modelfit light curves \citep[$0.71$\,mas/yr, see][]{0506wave}. Figure\,\ref{fig:waveplot} shows that the first total intensity flare coincides with the flare of the VLBI core. Moreover, the second total intensity flare occurs roughly at the same time when the stationary component 3 is flaring and the wave propagation is passing through it. As the wave propagates further downstream, it causes component 2 to flare, coinciding with the third peak in total intensity. In summary, the three sub-flares reported by \cite{Baikal0506} match the flare pattern of the core, component 3, and component 2, respectively. The effect that individual components can dominate the total intensity variability of a blazar has been observed before, e.g., in Fig.~2 of \cite{Valtaoja_1999} for the source 3C\,345, where an extended component even episodically outshines the core. However, it is remarkable that these flares can be linked to multiple components for the case of TXS\,0506+056, and that they relate to significantly faster motion than suggested by conventional jet kinematics, especially given their multimessenger counterpart.

\begin{figure}
    \centering
    \includegraphics[width=\linewidth]{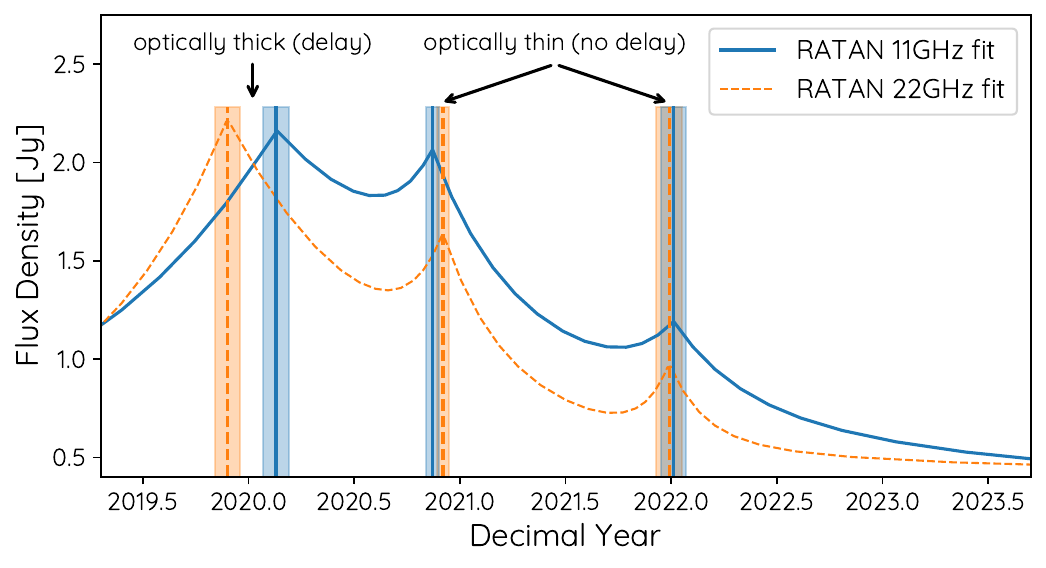}
    \caption{Single-dish light curve fits for TXS\,0506+056 at 22\,GHz (orange) and 11\,GHz (blue) based on RATAN-600 data, as published by \cite{Baikal0506}. The delay between the flare peaks can be an indicator whether the origin region of the emission is optically thick (delay) or optically thin (no delay).}
    \label{fig:ratan}
\end{figure}

We suggest that such geometrical flares can be revealed with single-dish multi-frequency radio monitoring data, even without knowledge of the VLBI structure of a source. Figure\,\ref{fig:ratan} shows the total intensity radio light curve fits at 11\,GHz and 22\,GHz of TXS\,0506+056, as published by \cite{Baikal0506} based on RATAN-600 observations. Here, it is clearly visible that the first flare has a time delay between the two frequencies, with the higher frequency peaking first, while the subsequent two flares happen simultaneously at both frequencies. The time delay in the first flare is typical for optically thick regions, such as the VLBI core, while simultaneous flares across frequencies could suggest an optically thin origin region, such as extended jet components \citep[e.g.,][]{Kutkin2019}. If similar multi-layered jet configurations with fast and slow moving jet plasma are present in other sources, this characteristic light curve pattern could serve as a valuable tracer to find shock-shock interactions in AGN jets, even without performing expensive VLBI observations.

We note that the short-lived, new-born component 4 was not detected in the total intensity light curve fits published by \cite{Baikal0506}. However, looking at their RATAN-600 data, there is a clear indication of an additional total intensity peak that would overlap with the flare of component 4.
Though, the exact nature of this newly born component is not entirely clear. It could be another stationary feature very close to the core that is lit up by the fast wave propagation that is otherwise too faint to be detected. 
Alternatively, in the scenario of an accelerated wave, component 4 could represent the initial (slower) wave shock that is later accelerated, possibly through interaction with the stationary features \citep[e.g.,][]{Fromm2013Kinematic}. Additionally, its ejection time suggests that it was likely produced close in time to the gamma-ray flare and neutrino event in 2017. Constraining the nature of component 4 better would only be possible with dense, higher-resolution, high-fidelity observations during the time of the wave propagation.

\subsection{Spine-sheath in the multimessenger context}

While hints of a spine-sheath scenario in TXS\,0506+056 have already been reported in previous VLBI observations by \cite{Ros2020},
The new polarization information presented in Fig.~\ref{fig:spine_sheath} provides compelling evidence for a stratified, spine–sheath jet in TXS\,0506+056, supporting conclusions of \cite{0506wave}. A spine-sheath model for TXS\,0506+056 was previously considered to explain the neutrino-emission related to the 2017 IceCube neutrino event by \cite{Ansoldi18}. In their model, neutrino emission is explained by the interaction of fast moving protons accelerated in the jet spine with an assumed Lorentz factor of $\Gamma_j=22$ that interact with external photons from a slower sheath layer with an assumed Lorentz factor of $\Gamma_s=2.2$, surrounding the spine. \cite{0506wave} have estimated the Lorentz factor of the fast spine, $\Gamma_\mathrm{spine} = 21$, as well as the jet viewing angle, $\theta=2.4^\circ$. Assuming the same viewing angle for the slower layers and considering the fastest of the modelfit components, i.e., component 1 with $\beta_\mathrm{app}=(1.93\pm0.45)$\,c as an upper limit for the speed of the slow layer (see Tab.\,\ref{tab:kinematic_fit}), the Lorentz factor of the slow jet layer can be constrained to $\Gamma_\mathrm{slow}\leq 5$. These Lorentz factor estimates are remarkably similar to the values required by the spine-sheath model presented by \cite{Ansoldi18}, i.e., $\Gamma_\mathrm{fast}=22$ and $\Gamma_\mathrm{slow}=2.2$. 

On the other hand, spine-sheath jets are widely discussed as a possible resolution of the Doppler crisis or bulk Lorentz factor crisis \citep[e.g.,][]{Tavecchio2014,Tavecchio2015}. This phenomenon describes the fact that TeV-emitting blazars, such as TXS\,0506+056, typically exhibit slow jet speeds and low brightness temperatures \citep[see Sect.\,\ref{sec:kinematics} and][]{PinerEdwards2018,Lister2019}, indicative of low Doppler factors. On the other hand, explaining their high-energy emission requires significantly higher Doppler factors, typically by an order of magnitude \citep[e.g.,][]{Tavecchio2010,Gokus2018}. While a spine-sheath jet configuration would naturally resolve this discrepancy, it has so far only been confirmed for very few bright TeV-emitting blazars, such as Mrk\,501 and Mrk\,421 by detection of limb-brightened jets in total intensity and characteristic polarization patterns \citep[e.g.,][]{Giroletti2004,Giroletti2008,PinerEdward2010}. Our results suggest that this apparent rarity may, at least in part, reflect observational limitations. In TXS\,0506+056, the spine-sheath morphology becomes apparent only through the combination of an exceptional broadband flare, which strongly enhances the polarized jet emission, and long-term, high-sensitivity VLBI polarization monitoring. This indicates that major flaring events can provide a unique opportunity to reveal the transverse jet structure in Doppler-crisis blazars that may otherwise remain undetected.

If this interpretation is correct, TXS\,0506+056 is unlikely to be a unique case. It is likely that similar features of stratified jets can be observed in other Doppler crisis blazars, especially in response to major multiwavelength flares. Therefore, this discovery opens up a new approach to systematically resolve the Doppler crisis. This is not only possible by tracing flux density waves in other sources with VLBI, as suggested by \cite{0506wave}, but also by searching for characteristic geometric flare features in multi-frequency light curves, as demonstrated in Fig.\,\ref{fig:ratan}. Moreover, high-sensitivity polarization studies of TeV-emitting AGN, especially during flaring periods, would be an ideal measure to search for possibly characteristic polarization patterns, similar to the one shown in Fig.\,\ref{fig:spine_sheath}. Radio monitoring programs that focus on TeV-emitting blazars, such as the TELAMON program \citep{TELAMON}, can play a key role in identifying such strong flares and performing dense follow-up observations at multiple frequencies to reveal subsequent flares from optically thin regions. This would ultimately present a scenario that could resolve the Doppler crisis for most, if not all, affected sources. 

\subsection{Comparison with previous VLBI studies}

Prior to the study presented here, several other works have aimed at probing the VLBI jet structure of TXS\,0506+056. However, the parsec-scale jet structure has remained inconclusive so far, especially following the major radio outburst in 2020. While some works suggest the presence of a dual-jet system \citep[e.g.,][]{Britzen2019} or a precessing, helical jet attributed to an underlying supermassive binary black hole system \citep[e.g.,][]{Li2020}, hints of a spine-sheath structure were previously reported by \cite{Ros2020}. In this work, we do not find compelling evidence for a binary black hole or dual-jet scenario, whereas the stacked polarization map in Fig.\,\ref{fig:spine_sheath} strongly supports the presence of a spine-sheath structure. Moreover, based on the stacked map, no ring-like features that could be caused by gravitational lensing, as reported by \cite{Britzen2025}, are evident.

The previous modelfit versions presented by \cite{Kun2019} and \cite{Li2020} agree with the presence of slow, quasi-stationary features that are also found in the kinematic analysis in Sect.\,\ref{sec:kinematics}. However, the wave propagation further discussed in the companion paper by \cite{0506wave} was not reported before, since it has become visible only after the major radio flare in 2020. Interestingly, the flux density wave propagation can also be seen in an independent modelfit analysis presented in Fig.\,1 of \cite{Song2025}, as well as Fig.\,5 of \cite{Li2020}, providing additional independent evidence for its existence.

In observations previous to the neutrino event, the jet of TXS\,0506+056 showed significant jet bending at larger scales beyond 2\,mas away from the core with the jet changing direction from south to south-east, as reported by \cite{Britzen2019} and \cite{Li2020}. As seen in Fig.\,\ref{fig:spine_sheath} (top left), the jet looks significantly more straightened than in previous works and a clear jet bend is not visible anymore at these scales. Therefore, the binary jet configuration suggested by \cite{Britzen2019} which claims the presence of a second VLBI-core $\sim1.5$\,mas south-west of the core component seems very unlikely.

\section{Summary and outlook} 
\label{sec:conclusions}

TXS\,0506+056 is one of the most tantalizing neutrino-candidate sources, given its strong multiwavelength outburst simultaneous with a high-energy neutrino detection in 2017 \citep{IceCubeTXS0506}.
In this work, we analyzed the dynamics of the parsec-scale jet of TXS\,0506+056 following the neutrino event, throughout a major radio flare that reached its peak in 2020. We performed a kinematic analysis based on 43 VLBI epochs observed with the VLBA as part of the MOJAVE program, observed between 2009 and 2025. Our visibility modelfits revealed the presence of several slow, quasi-stationary features. Additionally, we identified a newly born component with a moderate superluminal speed of $\sim(1-2)$\,c, whose ejection time is consistent with the neutrino event in 2017. \cite{0506wave} have revealed that the jet of TXS\,0506+056 is likely stratified, with significantly faster plasma propagating through a hidden ultrarelativistic spine. This causes the slower modelfit components, likely attributed to the sheath, to exhibit characteristic delayed flare peaks in their light curves. These component flares can episodically dominate the total intensity variability of the source, leading to characteristic geometric flares in multi-frequency light curves (see Fig.\,\ref{fig:waveplot} and Fig.\,\ref{fig:ratan}).

A polarimetric analysis of the modelfit components reveals that simultaneous to their total intensity flare, they exhibit a flare in linear polarization under significant rotation of their EVPA (see Fig.\,\ref{fig:pol_lcs}). We suggest that these rotations could arise from changing dominance between spine and sheath emission regions, leading to an apparent EVPA swing. The different intrinsic polarization of spine- and sheath layers is evident from the stacked polarization map compiled from all 43 VLBI epochs (see Fig.\,\ref{fig:spine_sheath}). We find a polarization pattern with EVPAs that are aligned with the jet in the center and perpendicular to the jet at its edges. A similar structure was previously observed in other AGN \citep[e.g.,][]{Pushkarev2005,Gabudzda2014} where it has been attributed to the a spine-sheath layered jet \citep{Tavecchio2014,Tavecchio2015}. 

For TXS\,0506+056 a spine-sheath scenario was previously considered to explain its neutrino-emission by \cite{Ansoldi18}. The required Lorentz factors for the slow and fast jet layers in their model ($\Gamma_\mathrm{slow}=2.2$; $\Gamma_\mathrm{fast}=22$) are remarkably similar to our estimates for the slow jet layer, $\Gamma_\mathrm{slow}\leq5$, and the fast spine from \cite{0506wave}, $ \Gamma_\mathrm{fast}=21$. The stratified jet also naturally resolves the Doppler crisis in TXS\,0506+056 by attributing the high Doppler factors required by high-energy studies to the fast spine, and the low Doppler factors suggested by previous radio observations to the sheath. We suggest that such hidden fast jet plasma could be revealed in other Doppler crisis blazars, especially in response to major flares. This can happen not only through direct VLBI observations \citep[e.g.,][]{0506wave} but also through detecting the associated characteristic geometric light curve flares, attributed to optically-thin jet features (see Fig.\,\ref{fig:waveplot} and \ref{fig:ratan}). Additionally, polarization imaging of Doppler crisis blazars offers another independent probe of characteristic stratified jet structure. 

While the stacking technique served as a powerful tool to reveal the spine-sheath structure in TXS\,0506+056, extracting additional information about its dynamics based on individual epochs during the wave propagation is significantly limited by the sensitivity of current-generation telescopes. With the superior sensitivity of future facilities, such as the SKA \citep{SKA} or the ngVLA \citep{ngVLA}, it might be possible to constrain the interplay of spine- and sheath-layers in full polarization to better understand the underlying physical processes. Moreover, future higher-resolution studies of TXS\,0506+056 with the Global Millimeter VLBI Array \citep[GMVA][]{GMVA} or the Event Horizon Telescope \citep[EHT, e.g., ][]{RoederEHT,JanssenCenA,Kim3C279} could provide valuable information about the nature of the stationary features and about the geometry near the jet base.

\begin{acknowledgements}

We thank Mischa Breuhaus for valuable comments that helped us improve the manuscript.
FE, MKa and ER acknowledge support from the Deutsche Forschungsgemeinschaft (DFG, grant 447572188, 443220636 [FOR5195: Relativistic Jets in Active Galaxies]). We acknowledge the M2FINDERS project from the European Research Council (ERC) under the European Union's Horizon 2020 research and innovation programme (grant agreement No~101018682). YYK and VAM were supported by the MuSES project, which has received funding from the European Union (ERC grant agreement No~101142396). Views and opinions expressed are those of the authors only and do not necessarily reflect those of the European Union or ERCEA. Neither the European Union nor the granting authority can be held responsible for them. ABP was supported in the framework of the State project ``Science'' by the Ministry of Science and Higher Education of the Russian Federation under the contract 075-15-2024-541. TS acknowledges support from the Research Council of Finland projects 362572 and 365088. This research has made use of the Astrophysics Data System, funded by NASA under Cooperative Agreement 80NSSC21M00561. The VLBA is an instrument of the National Radio Astronomy Observatory. The National Radio Astronomy Observatory and Green Bank Observatory are facilities of the U.S. National Science Foundation operated under cooperative agreement by Associated Universities, Inc.

This research has made use of data from the MOJAVE database that is maintained by the MOJAVE team \citep{MOJAVE}.

\end{acknowledgements}

\bibliographystyle{aa} 
\bibliography{0506kinematics} 

\begin{appendix}
\section{Evolution of Component Flux Dominance}
\label{app:additional_wave}

\begin{figure}
    \centering
    \includegraphics[width=\linewidth]{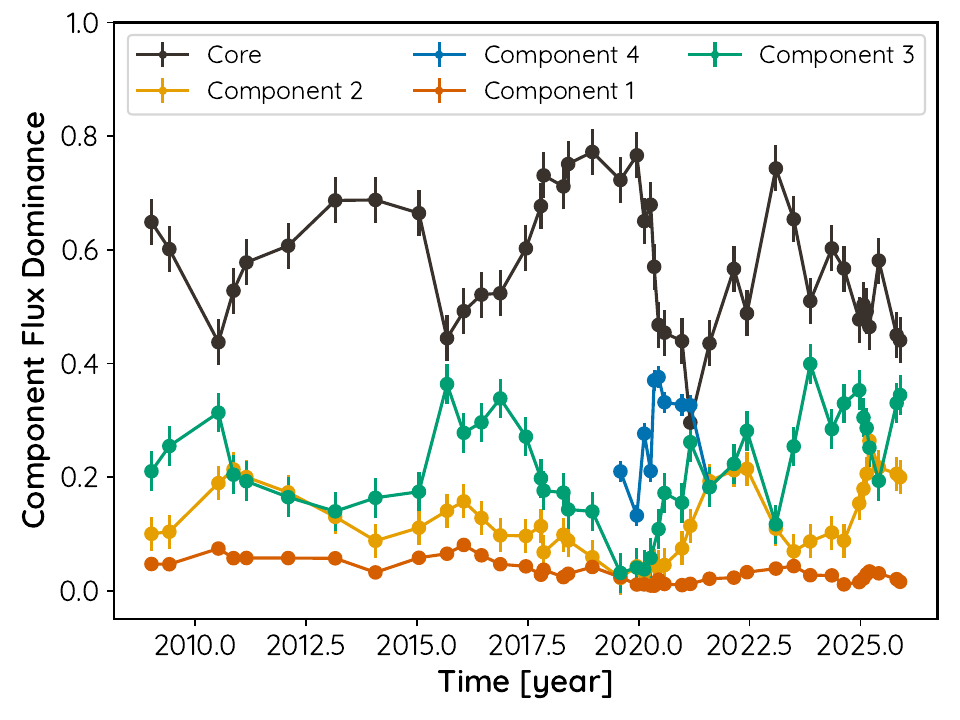}
    \caption{Component dominance (i.e., fraction of the total flux density) throughout the MOJAVE monitoring. The ultrarelativistic wave propagation discussed by \cite{0506wave} is clearly visible by a very high core dominance in 2019$-$2020, followed by a significant drop of the core dominance and a subsequent rise of the dominance of components 4, 3, and 2. A qualitatively similar pattern can be seen around 2009$-$2011 and 2015$-$16.}
    \label{fig:component_dominance}
\end{figure}

As reported by \cite{0506wave}, an ultra-relativistic spine in TXS\,0506+056 was revealed through the strong total intensity flare starting around 2020, with delayed flare peaks appearing progressively further downstream. Motivated by this behavior, we explored how the modelfit data presented in Sect.~\ref{sec:results} could be used to construct a simple diagnostic for identifying temporally ordered variability in VLBI jets, including in lower-flux-density regimes.

To investigate this, we normalized each VLBI epoch by its total flux density and analyzed the fractional flux density contributed by each model component as a function of time. The evolution of this fraction, referred to as the component dominance, is shown in Fig.\,\ref{fig:component_dominance} for all components. In this framework, the wave propagation revealed by \cite{0506wave} is characterized by a strong dominance of the core component around 2019$-$2020, reaching values close to 80\%. Shortly after, the core dominance drops significantly and the dominance of the extended components rises, one after the other. 

Qualitatively similar patterns are observed at other times in the dataset. A systematic assessment of the robustness, statistical significance, and physical interpretation of these additional features is challenging due to the lower flux density regime and corresponding increased uncertainties, and is therefore beyond the scope of this work. Nevertheless, the component-dominance representation introduced here may serve as a simple complementary diagnostic for highlighting similar features also in other, possibly fainter sources.

\end{appendix}

\end{document}